\documentclass[twocolumn,epjc3]{svjour3}          % twocolumn

\RequirePackage[T1]{fontenc}

\smartqed  % flush right qed marks, e.g. at end of proof

\RequirePackage{graphicx}
\RequirePackage{mathptmx}      % use Times fonts if available on your TeX system
\RequirePackage{flushend}
\RequirePackage[numbers,sort&compress]{natbib}
\RequirePackage[colorlinks,citecolor=blue,urlcolor=blue,linkcolor=blue]{hyperref}

\journalname{Eur. Phys. J. C}

\begin{document}

\title{A Jacobian elliptic single--field inflation}

\author{J. R. Villanueva\thanksref{e1,addr1,addr2}
        \and
        Emanuel Gallo\thanksref{e2,addr3,addr4} %etc.
}

%\thankstext[$\star$]{t1}{Thanks to the title}
\thankstext{e1}{e-mail: jose.villanuevalob@uv.cl}
\thankstext{e2}{e-mail: egallo@famaf.unc.edu.ar}

\institute{Instituto de F\'{\i}sica y Astronom\'ia, 
Universidad de Valpara\'iso, Gran Breta\~na 1111,
Valpara\'iso, Chile\label{addr1}
          \and
          Centro de Astrof\'isica de Valpara\'iso, Gran Breta\~na 1111, Playa Ancha,
Valpara\'iso, Chile\label{addr2}
          \and
          FaMAF, Universidad Nacional de C\'ordoba , 
Ciudad Universitaria, 5000 C\'ordoba, Argentina\label{addr3}
\and
Instituto de F\'isica
Enrique Gaviola (IFEG) CONICET Ciudad Universitaria, 5000 C\'ordoba, Argentina\label{addr4}
}

\date{Received: 30 March 2015 / Accepted: 18 May 2015}
% The correct dates will be entered by the editor

\maketitle

\abstract{
In the scenario of single-field inflation, this field
is described in terms of Jacobian elliptic functions. 
This approach provides, when constrained to particular 
cases, analytic solutions already known in the past, generalizing them
to a bigger family of analytical solutions. 
The emergent cosmology is analyzed using the 
Hamilton--Jacobi approach and then the main 
results are contrasted with the recent measurements 
obtained from the {\it Planck} 2015 data.
}

%\tableofcontents

%%%%%%%%%%%%%%%%%%%%%%%%%
\section{Introduction}\label{intro}
One of the main paradigms of the early universe 
is to the so-called {\it inflation}, which, for example, allows 
quantum fluctuations to produce the seeds for 
the cosmic microwave background anisotropy 
and large scale structure, so one of the ways by which this stage 
can emerge is, for example, protecting 
the holographic principle \cite{lobito}.
In the standard inflationary cosmology, 
the universe is dominated by a scalar inflaton field, $\phi$, 
self-interacting through a potential,
$V(\phi)$, with an energy density 
$\rho_{\phi}=\small{\frac{1}{2}} \dot{\phi}^2+V(\phi)$,
and pressure $p_{\phi}=\small{\frac{1}{2}} \dot{\phi}^2-V(\phi)$. 
Since inflation proceeds if the potential energy of the 
field dominates its kinetic energy, the weak energy condition 
ensures that the pressure is negative during the inflationary 
process allowing for an extremely rapid expansion of the universe.
In general, it is not easy to obtain exact solutions of the Einstein equations
with an inflaton field as source,
for a given potential $V(\phi)$.
However, for some simple cases, exact solutions can be obtained 
without even making use of the slow--roll approximation.

To perform a quantitative analysis of the inflation one could proceed
in two way: using the standard slow--roll approximation \cite{Liddle94},
or using the Hamilton--Jacobi method 
\cite{salopek,muslimov90,Kinney:1997ne,carr93}. In this work, 
we opt for the latter approach, 
obtaining new exact solutions for an inflaton field where the potential
is expressed in terms of elliptical functions. The reason behind this choice is the fact that many authors have reported potentials in terms of trigonometric \cite{barrow,chevron,Schunck,Kim}
and hyperbolic functions\cite{barrow,chevron,Schunck,chaadaev,hawkins01,delCampo:2012qb,delcampo13,Kim,harko}. 
Within this context, and keeping in mind that both, trigonometric and hyperbolic functions, are nothing but particular cases of elliptical functions, we seek for a generalization based on functions of this kind.

This paper is organized as follows: in Section II we give a review of the 
Hamilton--Jacobi approach to inflation. In Section III we present a quartic potential which 
can be expressed in terms of elliptical functions, and we provide the general solutions and the 
expressions for the relevant cosmological parameters. In Section IV we study scalar and tensor 
perturbations and we also compare them with the available observational
data.

\section{Hamilton-Jacobi approach to inflation}

The starting point is the expression of the equations 
governing the background
in terms of the inflaton $\phi$, so in units where 
$\hbar=c=1$ these relations are given by

\begin{equation}
\label{b1}
H^2=\frac{8\,\pi}{3\,m_p^2}\left[\frac{1}{2}\dot\phi^2 + V(\phi)\right]
\end{equation}
and
\begin{equation}
\label{b2}
\ddot{\phi}+3\,H\,\dot{\phi}+V'(\phi)=0,
\end{equation}
where $m_p=G^{-1/2}$ is the Planck mass, 
a dot represents differentiation with respect 
to the cosmic time, $t$, and a prime represents 
differentiation with respect to the inflaton scalar field.
Taking the temporal derivative of Eq. (\ref{b1}), we obtain
\begin{equation}
\label{b3}
H\,\dot{H}=\frac{4\,\pi}{3\,m_p^2}\,\left[\ddot{\phi}+V'(\phi) \right]\,\dot{\phi},
\end{equation}
so, using Eq. (\ref{b2}) into Eq. (\ref{b3}) leads to
the useful relation 
\begin{equation}
\label{b4}
\dot{\phi}^2=-\left(\frac{m^2_p}{4\,\pi}\right)\,\dot{H}.
\end{equation}
Also, considering the relation between operators
\begin{equation}
\label{b5}
\frac{d}{dt}=\dot{\phi}\,\frac{d}{d\phi},
\end{equation}
Equation (\ref{b4}) can be written as
\begin{equation}
\label{b6}
\dot{\phi}=-\left(\frac{m^2_p}{4\,\pi}\right)\,H'.
\end{equation}
If we assume that the Hubble parameter 
and the inflaton are invertible, then we can write
the potential as a function of $H$.
Therefore, inserting Eq. (\ref{b6}) into Eq. (\ref{b1})
we obtain the kinematic equation involving
the evolution of the scalar field
\begin{equation}
\label{b7}
\left(\frac{dH}{d\phi}\right)^2=\frac{12\,\pi}{m_p^2}\,H^2-\frac{32\pi^2}{m_p^4}\, V.
\end{equation}
Since the inflaton is a real scalar field, the condition
that $H^2>\frac{8\,\pi}{3\,m_p^2}\,V$ must be fulfilled during inflation.
Equations (\ref{b6}) and (\ref{b7}) are called the Hamilton--Jacobi
equations, and they dictate the kinematic evolution.
In order to obtain exact solutions to Eqs. (\ref{b6}) and  (\ref{b7}), we propose
a potential that leads to analytic solutions, i.e., that allows us
to find the generating function, $H(\phi)$.
Thus, the set of Eqs.  (\ref{b6}) and  (\ref{b7})
yield exact solutions. On the other hand,
from Eq. (\ref{b6}),  $a'\,H'=-\left(\frac{m^2_p}{4\,\pi}\right)\,a\,H$,
it follows that
\begin{equation}
\label{b8}
a(\phi)=a_i\,\exp\left[-\frac{4\,\pi}{m_p^2}\,\int_{\phi_i}^{\phi}
\frac{H(\phi)}{H'(\phi)} d\phi\right],
\end{equation}
where $a_i=a(\phi_i)$, and, from now on, the subscripts $i$ 
and $f$ imply that quantities are evaluated when inflation begins and ends, respectively.
The above equation implies that the scalar factor $a$
can be written as a function of the inflaton field $\phi$, and
thus, assuming that we know the scalar field as a function of time, 
we can obtain the scale factor as a function of the cosmological time.

It is also possible to express the acceleration equation
for the scale factor as
\begin{equation}\label{b9}
\frac{\ddot{a}}{a}=H^2 (1-\epsilon_H),
\end{equation}
where the  first Hubble hierarchy parameter  $\epsilon_H$ is given by
\begin{equation}\label{b10}
\epsilon_H\equiv - \frac{d \ln H}{d \ln a}=\left(\frac{m^2_p}{4\,\pi}\right)
\left(\frac{H'}{H}\right)^2.
\end{equation}
This parameter provides information as regards the acceleration of the universe, 
so during inflation the bound $\epsilon_H \ll 1$ is fulfilled, 
and inflation ends when $\epsilon_H=1$.

Another quantity to describe inflation is 
the number of e-folding of the physical expansion,
which is given by
\begin{equation}
\label{b11}
N\equiv \ln\left(\frac{a_f}{a_i}\right),
\end{equation}
so, using Eqs. (\ref{b6}), (\ref{b8}) and (\ref{b10}) we obtain
\begin{equation}
\label{b12}
N\equiv \int_{t_{i}}^{t_f} H dt=\left(\frac{4\,\pi}{m^2_p}\right)
\int_{\phi_f}^{\phi_i}\frac{H}{H'}d\phi=
\int_{\phi_f}^{\phi_i}\frac{1}{\epsilon_H}\frac{H'}{H}d\phi
\end{equation}

In the description of inflation it is also relevant to show 
that the solutions are independent from their initial 
conditions. This guarantees the true predictive power that  
any inflationary universe model must have, otherwise 
the corresponding physical quantities associated with 
the inflationary phase, such as the scalar or tensor 
spectra, would depend on these initial conditions. 
Thus, any inflationary model needs to fulfill the 
condition that its solutions follow an 
attractor behavior, in the sense that 
solutions with different initial conditions 
should converge to a unique solution \cite{salopek}.

Let us start by considering a linear perturbation, 
$\delta H(\phi)$, around a given inflationary solution, 
expressed by $H_{(0)}(\phi)$. In the following we will refer 
to this quantity as the background solution, and 
any quantity with the subscript $(0)$ is assumed to be evaluated taking into
account the background solution. 
Substituting $H=H_{(0)}+\delta H$ into
Eq. (\ref{b7}) and then linearizing, 
it is straightforward to show that \cite{Liddle94,guo03,guo04}
\begin{equation}
\label{b13}
\delta H \simeq \frac{m_p^2}{12\, \pi}\,\frac{H'_{(0)}}{H_{(0)}} \delta H',
\end{equation}
which can be solved as
\begin{equation}
\label{b14}
\delta H(\phi)=\delta H(\phi_i) \exp\left[ \int_{\phi_i}^{\phi}
\left(\frac{3}{\epsilon_H}\right)\,\frac{H'_{(0)}}{H_{(0)}} d\phi\right],
\end{equation}
where $\phi_i$ corresponds to some arbitrary initial value of $\phi$. 
Since $d\phi$ and $H'$ have opposite signs (assuming that $\dot{\phi}$
does not change sign due to the perturbation $\delta H$) the 
linear perturbations tend to vanish quickly \cite{Liddle94}.

\section{A quartic potential}

In this section we will present a class of potential 
which can be given in terms of elliptic functions. 
The reason behind this choice is the fact that many authors have reported potentials in terms of trigonometric \cite{barrow,chevron,Schunck,Kim}
and hyperbolic functions \cite{barrow,chevron,Schunck,chaadaev,hawkins01,delCampo:2012qb,delcampo13,Kim,harko}.  
So, recalling that both functions, trigonometric and hyperbolic, are 
special cases of elliptic functions, we will try to find 
a potential which leads to some function of this kind.
To proceed, we define the following potential: 
\begin{equation}
\label{p2.1}
\frac{8 \pi}{3 m_p^2}\,V=\alpha_0+ \alpha_1\,Y(H)
+\alpha_2\,Y^2(H)+\alpha_4\,Y^4(H),
\end{equation}
where the $\alpha_n$, for $n=1,2,3,4$,
are explicitly given by 
$\alpha_0=H_0^2-H_1^2(1-k^2)$,
$\alpha_1=2H_0 H_1$, $\alpha_2=2H_1^2(1-k^2)$,
$\alpha_4=H_1^2 k^2$, 
and the dimensionless function $Y$ is
given by
\begin{equation}
Y(H)=\frac{H}{H_1}-\mathcal{H}.
\end{equation}
In these definitions, 
$\mathcal{H}\equiv H_0/H_1$, $H_1$ and $k$ are constants which will be tested 
against observational data.

\begin{figure}
  \resizebox{0.47\textwidth}{!}{ \includegraphics{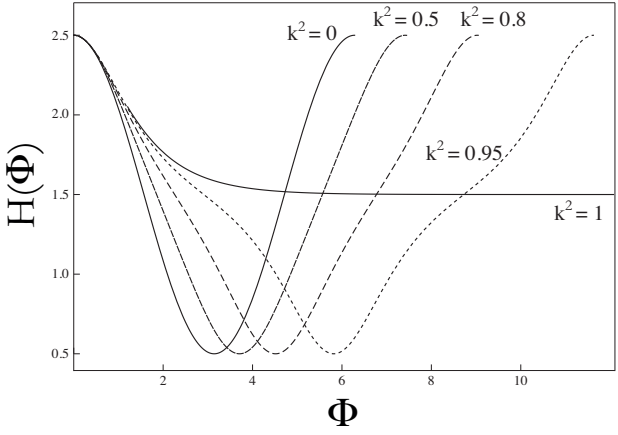}%
   }
  \caption{Plot of the Hubble parameter $H$ as a 
  function of the dimensionless scalar field $\Phi$.}
  \label{f1}
\end{figure}

\begin{figure}[h!]
  \resizebox{0.47\textwidth}{!}{ \includegraphics{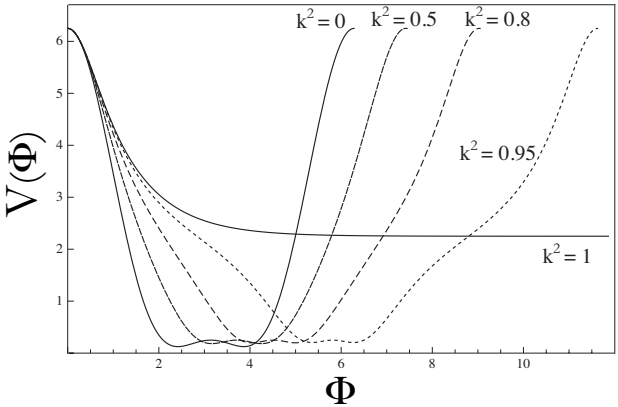}%
   }
  \caption{A characteristic behavior of the potential $V$ as a 
  function of the dimensionless scalar field $\Phi$ for different values
  of the modulus $k$. All plots were made using $\mathcal{H}=1.5$.}
  \label{f2}
\end{figure}

Substitution of Eq. (\ref{p2.1}) into the general Eq.
(\ref{b7}) implies that
\begin{equation}
\label{p2.2}
\left(\frac{1}{\beta}\frac{dY}{d\phi} \right)^2=
(1-Y^2)(k'^2+k^2\,Y^2),
\end{equation}
where $\beta^2=12\pi/m_p^2$, and $k'^2=1-k^2$.
Therefore, performing an integration (see \cite{byrd,Armitage,Mey0l,hancock,tablas}), 
and then solving for $H$,
we obtain 
\begin{equation}
\label{p2.3}
H(\phi)=H_1\{ \mathcal{H} +\,\textrm{cn}\left[\beta\,(\phi-\phi_0)\right]\},
\end{equation}
where $\textrm{cn}(x)\equiv\textrm{cn}(x|\, k)$ is the Jacobi elliptic cosine function.
Fig.\ref{f1} depicts $H(\phi)$ for different values
of the {\it modulus} $k$.
So, by defining the dimensionless scalar field  $\Phi=\beta\,(\phi-\phi_0)$, 
and  after a slight manipulation, the potential is 
conveniently written as
\begin{equation}
\label{p2.4}
\mathcal{V}(\phi)\equiv \frac{8 \pi}{3 m_p^2}\,\frac{V}{H_1}=(\mathcal{H}+\textrm{cn}\,\Phi)^2 -
\textrm{sn}^2\,\Phi\,
\textrm{dn}^2\,\Phi,
\end{equation}
where $\textrm{sn}(x)\equiv\textrm{sn}(x|\,k)$ is the Jacobi elliptic sine function, and
$\textrm{dn}(x)\equiv\textrm{dn}(x|\,k)$ is
the Jacobi elliptic delta function.
The behavior of $V(\phi)$ is indicated in Fig. \ref{f2}, 
where for comparison, different values of the modulus $k$ are shown.
The dimensionless pressure and the energy density are given by
\begin{equation}
\label{dens}
\overline{\rho}_{\phi}\equiv \frac{8 \pi}{3 m_p^2}\,\frac{\rho_{\phi}}{H_1^2}=(\mathcal{H}+ \textrm{cn}\,\Phi)^2
\end{equation}
and
\begin{equation}
\label{press}
\overline{p}_{\phi}\equiv \frac{8 \pi}{3 m_p^2}\, \frac{p_{\phi}}{H_1^2}=-(\mathcal{H}+\textrm{cn}\,\Phi)^2+\textrm{sn}^2\,\Phi \,
\textrm{dn}^2\,\Phi,
\end{equation}
respectively. Therefore, the parameter of the equation of state, $\omega_{\phi}(\rho_{\phi})=p_{\phi}/\rho_{\phi}$, can be written as
\begin{equation}
\label{eos1}
\omega_{\phi}(\rho_{\phi})=-1+\sqrt{f_{\phi}},
\end{equation}
where the function $f_{\phi}\equiv f(\rho_{\phi})$ is given explicitly by
\begin{equation}
\label{ff1}
\frac{(1-\mathcal{H}^2+2\mathcal{H}\,\rho^{1/2}-\rho)[1-k^2(1-\mathcal{H}^2+2\mathcal{H}\,\rho^{1/2}-\rho)]}{\rho^2}.
\end{equation}
Note that the condition $f_{\phi}\geq 0$ must be satisfied for every
pair ($\mathcal{H}, k$). 

\begin{figure}
  \resizebox{0.47\textwidth}{!}{ \includegraphics{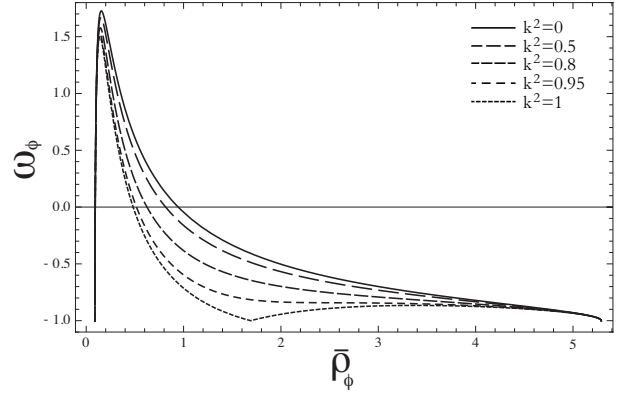}%
   }
  \caption{Plot of the parameter of the equation of state $\omega_{\phi}$ as a 
  function of the dimensionless density $\overline{\rho}_\phi$ for different values
  of the modulus $k$. All plots were made using $\mathcal{H}=1.3$.}
  \label{ome}
\end{figure}
Also, from Eq. (\ref{eos1}) we can see
that the Jacobian scalar field $\phi$
is driving inflation
in such way that the condition
$\omega_{\phi}\leq -1$ always is satisfied. Clearly
there are four critical values for the (dimensionless) density which 
lead to a cosmological-like constant regime ($\omega_{\phi}=-1$):
\begin{eqnarray}
\label{dens}
\overline{\rho}_{\phi,1, 2}^{(c)}&=&(\mathcal{H}\pm 1)^2,\\
\overline{\rho}_{\phi,3, 4}^{(c)}&=&\mathcal{H}^2+\left(1-\frac{1}{k^2}\right)\pm 2
\sqrt{\mathcal{H}^2 \left(1-\frac{1}{k^2}\right)}.
\end{eqnarray}
First of all, $\overline{\rho}_{\phi,1}^{(c)}$ and $\overline{\rho}_{\phi,2}^{(c)}$ 
are positive and independent of the modulus $k$. 
But, since $1\geq k \geq 0$, then $\overline{\rho}_{\phi,3}^{(c)}$ 
and $\overline{\rho}_{\phi,4}^{(c)}$  
are complex (a conjugate pair) except in the case $k = 1$, 
where they take the value (degenerate) 
$\overline{\rho}_{\phi,1}^{(c)}=\overline{\rho}_{\phi,2}^{(c)}=\mathcal{H}^2$,
see Fig. \ref{ome}.

Using the generating function (\ref{p2.3}) in Eq. (\ref{b8}), and then
performing an integration,
we find
\begin{equation}
\label{p2.5}
\frac{a(\phi)}{a_i}=\frac{\textrm{sd}\,\Phi}{\textrm{sd}\,\Phi_i}
\left(
\frac{\mathcal{G}(\phi)}{\mathcal{G}(\phi_i)}\right)^{\mathcal{H}},
\end{equation}
where $\textrm{sd}(x | k)=\textrm{sn}(x | k) / \textrm{dn}(x | k)$, and the Gugu 
function $\mathcal{G}(\phi)$ is given by
\begin{equation}
\mathcal{G}(\phi)=\frac{1+\textrm{cn}\,\Phi}{\textrm{sn}\,\Phi}\,
e^{\frac{k}{k'} \arctan\left(\frac{k}{k'}\,\textrm{cn}\,\Phi\right)}.
\end{equation}
Since this function diverges when $k=1$ (or $ k'=0$)
we must exclude it and integrate directly this case from Eq. (\ref{b8}),
obtaining
\begin{equation}
\label{afi2}
\frac{a(\phi)}{a_i}=\frac{\sinh\Phi}{\sinh\Phi_i}
\left(
\frac{\mathcal{F}(\phi)}{\mathcal{F}(\phi_i)}\right)^{\frac{H_0}{H_1}},
\end{equation}
with the function $\mathcal{F}(\phi)$ defined by
\begin{equation}
\label{ff}
\mathcal{F}(\phi)=\frac{\cosh\Phi-1}{\sinh\Phi}\,e^{\cosh\Phi}.
\end{equation}
In the same way, using the fact that $H'(\phi)=-\beta\,H_1\,\textrm{sn}\,\Phi \,
\textrm{dn}\,\Phi$ together with Eq. (\ref{b6}),
we see that the cosmological time is $t=\ln \mathcal{G}^{\frac{1}{H_1}}(\phi)$,
or more conveniently
\begin{equation}
\label{at}
\mathcal{G}(\phi)=e^{H_1 t},
\end{equation}
if $0\leq k <1$, whereas in the hyperbolic 
limit, $k=1$, the cosmological time is $t=\ln \mathcal{F}^{\frac{1}{H_1}}(\phi)$, or
\begin{equation}
\label{ath}
\mathcal{F}(\phi)=e^{H_1 t}.
\end{equation}

Note from Eqs. (\ref{p2.5}), (\ref{at}) and (\ref{ath}) 
we note that
$a \propto e^{H_0 (t-t_i)}$, required by the
de Sitter stage.

In addition, making use of the properties of the Jacobian elliptic functions,
we can write the first Hubble hierarchy parameter (\ref{b10})
in the convenient form
\begin{equation}
\label{p2.7}
\epsilon_H=\frac{
\textrm{sn}^2\, \Phi\, \textrm{dn}^2\, \Phi}
{[\mathcal{H} +\textrm{cn}\,\Phi]^2}.
\end{equation}
In the limit $k\rightarrow 0$, this simplifies to
\begin{equation}
\label{p2.8}
\epsilon_H=\frac{
\sin^2\,\Phi}
{[\mathcal{H} +\cos\,\Phi]^2},
\end{equation}
while in the limit $k\rightarrow 1$, we obtain
\begin{equation}
\label{p2.9}
\epsilon_H=\frac{
\textrm{sech}^2\,\Phi\,\textrm{tanh}^2\Phi}
{[\mathcal{H} +\textrm{sech}\,\Phi]^2}.
\end{equation}
Let us stop briefly in this last expression for the case $\mathcal{H}=H_0=0$,
which, except for a constant, leads to the 
expression obtained by del Campo \cite{delcampo13}
when a  Chaplygin-like scalar field is considered.

For the case of inflation ($\ddot{a}/a>0$), 
the relevant boundary condition is de Sitter expansion,
$\epsilon_H(\phi_S)=0$. So, for our case this condition
leads to $\phi_S=\phi_0$.
From the definition of $\epsilon_H$, 
this is equivalent to $H'(\phi)=0$, and it
follows that $\phi_0$ is a stationary point of the field:
\begin{equation}
\label{p2.11}
\dot{\phi}_{ |_{\phi=\phi_0}}=-\frac{m_p^2}{4 \pi}\,H'(\phi_0)=0.
\end{equation} 
From the generating function (\ref{p2.3}) we see that this point
is $\phi_0=0$. Note that at this point the Hubble parameter 
is given by $H(\phi_0)=H_0+H_1$.

\begin{figure}
  \resizebox{0.47\textwidth}{!}{ \includegraphics{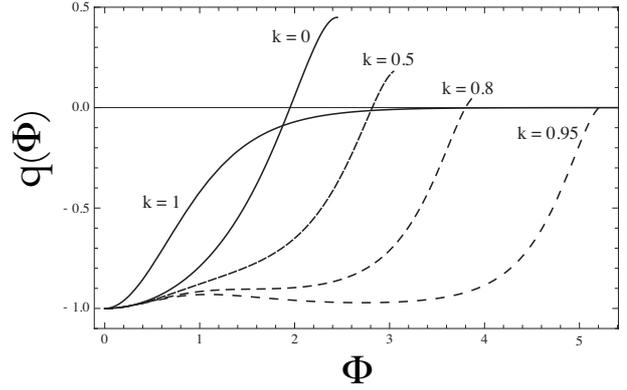}%
   }
  \caption{Plot of the deceleration parameter $q$ as a 
  function of the dimensionless scalar field $\Phi$.}
  \label{f3}
\end{figure}

In order to assess  whether we have an inflationary period, 
we introduce the deceleration parameter $q$, defined 
as $q = -\frac{\ddot{a} a}{\dot{a}^2}$, which, in terms of  the first Hubble hierarchy 
parameter  $\epsilon_H$, it becomes 
$q =\epsilon_H-1$, which gives
\begin{equation}
q=-1+\frac{
\textrm{sn}^2\, \Phi\, \textrm{dn}^2\, \Phi}
{[\mathcal{H} +\textrm{cn}\,\Phi]^2}.
\end{equation}
The behavior of $q(\phi)$ is indicated in Fig. \ref{f3} 
for different values of the modulus $k$.

The number of the e-folds is found to be
\begin{equation}\label{p2.12}
N(\Phi_i)=\ln\left(\frac{\textrm{sd}\,\Phi_f}{\textrm{sd}\,\Phi_i}\right)+
\mathcal{H}\ln\left(\frac{\mathcal{G}(\Phi_f)}{\mathcal{G}(\Phi_i)}\right).
\end{equation}

We must remark that the final value of the dimensionless 
scalar field $\Phi_f$ is obtained using the transcendental
equation
\begin{equation}
\label{teff}
\mathcal{H} =-\textrm{cn}\,\Phi_f\pm \textrm{sn}\, \Phi_f\, \textrm{dn}\, \Phi_f.
\end{equation}
From Eqs. (\ref{p2.12})
and (\ref{teff}) using $N=60$, $N=70$, $\mathcal{H}=1.3$ 
(except for $k=1$ where we have used $\mathcal{H}=0$)
and different values of the modulus $k$, some values of $(\phi_i, \phi_f)$
are presented in table \ref{tab:1}.

\begin{table*}
\caption{Values for the dimensionless scalar field
for some values of the modulus $k$. The values of $\Phi_f$
were calculated using the condition $\epsilon_H(\Phi_f)=1$ from
Eq. (\ref{p2.7}), while the values of $\Phi_i$ were calculated 
using the condition $N(\Phi_i)=60$ and $N(\Phi_i)=70$ 
from Eq. (\ref{p2.12}). In all these computations we use 
$\mathcal{H}=1.3$, except for $k=1$ where we have assumed
$\mathcal{H}=0$.}
\label{tab:1}
\begin{tabular*}{\textwidth}{@{\extracolsep{\fill}}lrrrrl@{}}
\hline
\multicolumn{2}{|c|}{\,}& \multicolumn{2}{|c|}{$N=60$}& \multicolumn{2}{|c|}{$N=70$}\\
\hline
$k^2$&$\Phi_f$&$\Phi_i$& $\Delta \Phi$ &$\Phi_i$& $\Delta \Phi$\\
\hline
0&1.9515&1.5764&0.3752 &1.5757&0.3760\\
\hline
0.5&2.8170&1.6883&1.1287&1.6863&1.1307\\
\hline
0.8&4.0568&2.0118&2.045&2.0088&2.0480\\
\hline
0.95&5.3123&2.6781&2.6342&2.6747&2.6376\\
\hline
1&19.5933& 1.1206&18.4728&1.0845&18.5088\\
\hline
\end{tabular*}
\end{table*}

The study of the attractor behavior of the model is performed
by considering a linear perturbation, $\delta H(\phi)$,
around  a given solution and following the steps described above.
Thus, from Eqs. (\ref{b14}), (\ref{p2.3}), and (\ref{p2.7}), we get that
\begin{equation}
\label{p2.13}
\delta H(\phi)=\left[\frac{\textrm{sd}\,\Phi_i}{\textrm{sd}\,\Phi}
\left(
\frac{\mathcal{G}(\phi_i)}{\mathcal{G}(\phi)}\right)^{\mathcal{H}}\right]^3 \delta H(\phi_i).
\end{equation}
Since the factor accompanying 
$\delta H(\phi_i)$ is nothing but $(a/a_i)^{-3}$
[see the expression (\ref{p2.5})], which during inflation increases at least
70 e-fold, we can conclude that $\delta H(\phi)$ is very small.

\section{Scalar and tensor perturbations}
In the inflationary scenario, the quantum fluctuations 
are found to be 
relevant because they generate two important types of perturbations: 
density perturbations (arising from quantum fluctuations 
in the scalar field, together with the corresponding scalar metric 
perturbation \cite{Lukash:1980iv,Hawking:1982cz,Starobinsky:1982ee}), 
and relic gravitational waves (which are tensor metric fluctuations 
\cite{Grishchuk:1974ny,Starobinsky:1979ty,Rubakov:1982df,Fabbri:1983us,Abbott:1984fp}). 
The former is sensitive to gravitational instability and leads to 
structure formation \cite{Mukhanov:1990me}, while the 
latter predicts a stochastic background of relic gravitational 
waves which could influence the cosmic microwave 
background anisotropy via the presence of polarization 
in it \cite{Kamionkowski:1996zd,Knox:2002pe}. 

In order to describe these  perturbations 
we  introduce a series of parameters known as {\it the Hubble hierarchy parameters}. 
We have already defined one of them through  Eq. (\ref{b10}), presented as 
the first Hubble hierarchy parameter. {\it The second Hubble hierarchy parameter}, 
$\eta_H$, is defined by

\begin{equation}
\label{eta}
\eta_H\equiv -\frac{d\, \ln H'}{d\,\ln a}=
\left(\frac{m_p^2}{4 \pi}\,\right)\,\left(\frac{H''}{H}\right),
\end{equation}
where plugging in the Jacobian scalar field leads to
\begin{equation}
\label{etaJ}
\eta_H=\frac{\textrm{cn}\,\Phi \,(1-2 \,\textrm{dn}^2\,\Phi)}{\mathcal{H}+
\textrm{cn}\,\Phi}.
\end{equation}
{\it The third Hubble hierarchy parameter}, $\xi_H^2$ is defined by
\begin{equation}
\label{xi}
\xi_H^2\equiv \left(\frac{m_p^2}{4 \pi}\,\right)^2\,\left(\frac{H'''\,H'}{H^2}\right),
\end{equation}
leading to
\begin{equation}
\label{xiJ}
\xi_H^2=\frac{\textrm{sn}^2\,\Phi\,\textrm{dn}^2\,\Phi
\,(6 k^2\,\textrm{sn}^2\,\Phi-4k^2-1)}
{[\mathcal{H}+
\textrm{cn}\,\Phi]^2}.
\end{equation}

The quantum fluctuations produce a power spectrum of scalar density fluctuations
of the form \cite{Guth:1982ec,Bardeen:1983qw}
\begin{equation}
\label{sdf}
\mathcal{P}_{\mathcal{R}}(\widehat{ \rm{k}})=\left(\frac{H}{|\dot{\phi}|}\right)^2
\left(\frac{H}{2\pi}\right)^2 \bigg\vert_{aH=\widehat{ \rm{k}}}.
\end{equation}
This perturbation is evaluated when a given mode $\widehat{ \rm{k}}$
crosses outside the horizon during inflation, i.e.
at $a H=\widehat{ \rm{k}}$. These modes do not evolve
outside the horizon, so we can assume they keep a
fixed value after crossing the horizon during
inflation.
In order to obtain some comparison with 
the available observational data, we introduce
the scalar spectral index $n_s$ defined as
\begin{equation}
\label{ns}
n_s-1\equiv \frac{d \ln \mathcal{P}_{\mathcal{R}}}{d \ln \widehat{ \rm{k}}}.
\end{equation}
After a brief calculation, one obtains that
\begin{equation}
\label{ns}
n_s=1 -4 \epsilon_H+2\eta_H,
\end{equation}
so, using Eqs. (\ref{p2.7}) and (\ref{etaJ}), the spectral index becomes
%\begin{widetext}

\begin{equation}
\resizebox{.9\hsize}{!}{$n_s=1+\frac{2 \left\{\mathcal{H}\,  \textrm{cn}\,\Phi (1-2\textrm{dn}^2\,\Phi)
+ \left[\textrm{cn}^2\,\Phi-3\textrm{dn}^2\,\Phi \left(\frac{5}{3}\textrm{cn}^2\,\Phi-1\right)\right] \right\}}{[\mathcal{H}+\textrm{cn}\,\Phi]^2}.$}
\label{siJ}
\end{equation}

%\end{widetext}

\begin{figure}
  \resizebox{0.47\textwidth}{!}{ \includegraphics{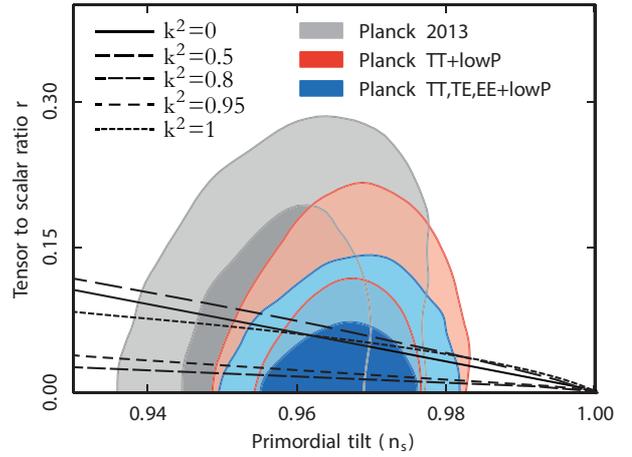}%
   }
  \caption{This plot shows the parameter $r$ as a function of the scalar spectral index 
  $n_s$ for various values of the modulus $k$, i.e. $k=0$, $k^2=0.5$,
  $k^2=0.8$, $k^2=0.95$ and $k=1$. The marginalized joint 68\% and 95\% confidence level regions using Planck TT+ lowP or Planck TT, TE, EE + lowP are shown \cite{planck15}. Also, constraints from the Planck 2013 data release are also shown for comparison. Every curves have been plotted with
$H_1=1$, $\mathcal{H}=1.3$, and they are normalized in such way that $r=0$
  when $n_s=1$.}
  \label{fr}
\end{figure}

It is well known, that not only scalar curvature perturbations are generated during inflation.
In addition quantum
fluctuations generate transverse-traceless tensor perturbations \cite{Mukhanov:1990me}, 
which do not couple to matter. Therefore they are only determined by the dynamics of the background metric. 
The two independent polarizations evolve like minimally coupled massless fields with spectrum
\begin{equation}
\label{PT}
\mathcal{P}_{\mathcal{T}}=\frac{16 \pi}{m_p^2} \left(\frac{H}{2\pi}\right)^2
\bigg\vert_{aH=\widehat{ \rm{k}}}.
\end{equation}
In the same way as the scalar perturbations, it is possible to introduce the gravitational wave 
spectral index $n_T$ defined by 
$n_T\equiv \frac{d \ln \mathcal{P}_{\mathcal{T}}}{d \ln \widehat{ \rm{k}}}$, 
which in our case becomes $n_T = -2\epsilon_H$ . At this 
point, we can introduce the tensor-to-scalar amplitude ratio 
$r \equiv \frac{\mathcal{P}_{\mathcal{T}}}{\mathcal{P}_{\mathcal{R}}}$
which becomes
\begin{equation}
\label{r}
r=4\epsilon_H,
\end{equation}
and, in our case, we obtain that
\begin{equation}
\label{rJ}
r= \frac{4\,
\textrm{sn}^2\, \Phi\, \textrm{dn}^2\, \Phi}
{[\mathcal{H} +\textrm{cn}\,\Phi]^2}.
\end{equation}
In Fig. \ref{fr} we display the parameter space of $r$, given by Eq. (\ref{rJ}),
and the spectral index given by Eq. (\ref{siJ}).

\section{Final remarks}
There is no doubt that elliptic functions are a powerful 
tool  to describe natural phenomena. 
In this paper we have used them to describe a scalar 
field in order to find inflationary solutions.
In particular, we have used an inflationary universe model 
in which the inflaton field is characterized
by a parameter of the equation of state 
$\omega_{\phi} = -1+\sqrt{f_{\phi}}$, where 
$f_{\phi}$ is given by Eq. (\ref{ff1}), and its kinematical
evolution was described by the Hubble parameter  
given by $H(\phi)=H_1\{ \mathcal{H} +\,\textrm{cn}\left[\beta\,(\phi-\phi_0)\right]\}$. 
A detailed analysis shows that the allowed values
for $H(\phi)$ depend strongly on the values
of the pair ($k, \mathcal{H}$).
On this basis, the scalar potential,
$V(\phi)$,
the corresponding number of e-foldings, 
and the attractor feature of the model 
were described. 

We should mention here that the first hierarchy parameter $\epsilon_H$
(and thereby, the deceleration parameter $q$)
related to the inflaton field results in such a way that it is possible to reproduce a generalized Chaplygin gas by making $\mathcal{H}\rightarrow 0$.
This analogy encouraged us to start a generalization 
of the GCG in terms of elliptic functions
\cite{vill15}.

It was also shown that the potential works quite 
well when compared against the measurements 
recently released based on Planck data. This situation 
is the main motivation to study inflationary universe 
models with this kind of scalar potential.

Finally, in general terms, we have found that the 
tensor-to-scalar ratio can adequately accommodate the currently available observational data for some values of the ($\mathcal{H}, k$) 
parameters. 
In this context, we have shown that the model here presented 
is appropriate for describing inflationary universe models.

%%%%%%%%%%%%%%%%%%%%%%%%%%%%%%%%%%%%%%%%%%%%%%%%%%%%%%%%%%%%%
\appendix

\section{ A brief review of Jacobian elliptic functions}\label{app:jef}
As a starting point, let us consider
the elliptic integral
\begin{eqnarray}\nonumber
u(y, k)&\equiv& u=\int_0^y\frac{dt}{\sqrt{(1-t^2)(1-k^2\,t^2)}}\\\label{a1}
&=&
\int_0^{\varphi}\frac{d\theta}{\sqrt{1-k^2\sin^2\theta}}=
F(\varphi, k),
\end{eqnarray}
where $F(\varphi, k)$ is the {\it normal elliptic integral of the first kind},
and $k$ is the {\it modulus}.
The problem of the inversion of this integral was
studied and solved by Abel and Jacobi, and this leads
to the inverse function defined by
$y=\sin\varphi=\textrm{sn}(u, k)$
with $\varphi=\textrm{am}\, u$, and 
the functions are called
{\it Jacobi elliptic sine} $u$ and {\it amplitude} $u$.

The function sn $u$ is an odd elliptic function of order two. It possesses
a simple pole of residue $1/k$ at every point congruent to
$i K'$ (mod $4K$, 2$i K'$) and a simple pole of residue
$-1/k$ at points congruent to $2K+i K'$ (mod 4 $K$, $2 i K'$),
where $K\equiv K(k)=F(\pi/2, k)$ is the {\it complete elliptic integral of the first kind},
$K'=F(\pi/2, k')$, and $k'=\sqrt{1-k^2}$ is the {\it complementary modulus}.

Two other functions can then be defined by 
$\textrm{cn}(u, k)\,\,\,=\,\,\,\sqrt{1-y^2}\,\,\,=\,\,\,\cos \varphi$, \quad
$\textrm{dn}(u, k) \,\,\,= \,\,\,\sqrt{1-k^2\,y^2} \,\,\,= \,\,\,\Delta \varphi \, \, \,=$
\newline 
$\sqrt{1-k^2\,\sin \varphi}$.
The set of functions $\{\textrm{sn}\, u, \textrm{cn}\,u, \textrm{dn}\, u\}$
are called {\it Jacobian elliptic functions}, and they
take the following special values:
\begin{eqnarray}
&&\textrm{sn} (u| 0)=\sin u,\quad \textrm{sn} (u| 1)=\tanh u,\\
&&\textrm{cn} (u| 0)=\cos u,  \quad \textrm{cn} (u| 1)=\textrm{sech}\, u,\\
&& \textrm{dn} (u| 0)=1,\, \qquad \,\,\textrm{dn} (u| 1)=\textrm{sech}\, u.
\end{eqnarray}
Some fundamental relations between Jacobian elliptic functions
are 
\begin{eqnarray}
&&\textrm{sn}^2 u +  \textrm{cn}^2 u=1,\\
&&k^2 \textrm{sn}^2 u + \textrm{dn}^2 u=1,\\
&& \textrm{dn}^2 u-k^2\textrm{cn}^2 u=k'^2,\\
&& k'^2\textrm{sn}^2 u +  \textrm{cn}^2 u=\textrm{dn}^2 u.
\end{eqnarray}

%%%%%%%%%%%%%%%%%%%%%%%%%%%%%5
\begin{acknowledgement}
J. R. Villanueva dedicates this paper to his master and friend, Sergio del Campo. 
E. Gallo is thankful for kind hospitality at Instituto de F\'isica y
Astronom\'ia, Universidad de Valpara\'iso,  while working on this paper. 
We would like to thank Amelia Bayo, V\'ictor C\'ardenas, Ram\'on Herrera, Osvaldo Herrera, 
and Cuauhtemoc Campuzano
for helpful conversations relating to this work.
E. Gallo acknowledges financial support from CONICET and SeCyT-UNC.
J. R. Villanueva is supported by Comisi\'on Nacional de Investigaci\'on Cient\'ifica 
y Tecnol\'ogica through FONDECYT Grants No. 11130695.
\end{acknowledgement}

%\nocite{*}
%\bibliographystyle{spr-mp-nameyear-cnd}
%\bibliography{myref}
%\bibliography{biblio-u1}

\end{document}